\documentclass[12pt]{article}
\usepackage{amssymb}

\def\hybrid{\topmargin 0pt      \oddsidemargin 0pt
        \headheight 0pt \headsep 0pt
        \voffset=-0.5cm
        \hoffset=-0.25in
        \textwidth 6.75in
        \textheight 9.5in       
        \marginparwidth 0.0in
        \parskip 5pt plus 1pt   \jot = 1.5ex}
\catcode`\@=11
\def\marginnote#1{}

\newcount\hour
\newcount\minute
\newtoks\amorpm
\hour=\time\divide\hour by60 \minute=\time{\multiply\hour by60
\global\advance\minute by-\hour}
\edef\standardtime{{\ifnum\hour<12 \global\amorpm={am}%
        \else\global\amorpm={pm}\advance\hour by-12 \fi
        \ifnum\hour=0 \hour=12 \fi
        \number\hour:\ifnum\minute<10 0\fi\number\minute\the\amorpm}}
\edef\militarytime{\number\hour:\ifnum\minute<10 0\fi\number\minute}

\def\draftlabel#1{{\@bsphack\if@filesw {\let\thepage\relax
   \xdef\@gtempa{\write\@auxout{\string
      \newlabel{#1}{{\@currentlabel}{\thepage}}}}}\@gtempa
   \if@nobreak \ifvmode\nobreak\fi\fi\fi\@esphack}
        \gdef\@eqnlabel{#1}}
\def\@eqnlabel{}
\def\@vacuum{}
\def\draftmarginnote#1{\marginpar{\raggedright\scriptsize\tt#1}}
\def\draftlabel#1{{\@bsphack\if@filesw {\let\thepage\relax
   \xdef\@gtempa{\write\@auxout{\string
      \newlabel{#1}{{\@currentlabel}{\thepage}}}}}\@gtempa
   \if@nobreak \ifvmode\nobreak\fi\fi\fi\@esphack}
        \gdef\@eqnlabel{#1}}
\def\@eqnlabel{}
\def\@vacuum{}
\def\draftmarginnote#1{\marginpar{\raggedright\scriptsize\tt#1}}

\def\draft{\oddsidemargin -.5truein
        \def\@oddfoot{\sl preliminary draft \hfil
        \rm\thepage\hfil\sl\today\quad\militarytime}
        \let\@evenfoot\@oddfoot \overfullrule 3pt
        \let\label=\draftlabel
        \let\marginnote=\draftmarginnote
   \def\@eqnnum{(\theequation)\rlap{\kern\marginparsep\tt\@eqnlabel}%
\global\let\@eqnlabel\@vacuum}  }

\def\numberbysection{\@addtoreset{equation}{section}
        \def\theequation{\thesection.\arabic{equation}}}

\def\underline#1{\relax\ifmmode\@@underline#1\else
        $\@@underline{\hbox{#1}}$\relax\fi}

\def\titlepage{\@restonecolfalse\if@twocolumn\@restonecoltrue\onecolumn
     \else \newpage \fi \thispagestyle{empty}\c@page\z@
        \def\thefootnote{\fnsymbol{footnote}} }

\def\endtitlepage{\if@restonecol\twocolumn \else  \fi
        \def\thefootnote{\arabic{footnote}}
        \setcounter{footnote}{0}}  

\numberbysection \hybrid

\newcounter{mo}

\newcommand{\tr}{{\rm tr}}
\newcommand{\ti}[1]{\tilde{#1}}

\newcommand{\vf}{\varphi}
\newcommand{\al}{\alpha}
\newcommand{\be}{\beta}

\newcommand{\om}{\omega}
\newcommand{\vth}{\vartheta}

\newcommand{\Mat}{ {\rm Mat}(N,\mathbb C) }

\newcommand{\mC}{\mathbb C}

\def\beq{\begin{equation}}
\def\eq{\end{equation}}
\def\p{\partial}

\begin{document}

\setcounter{page}{1}

\date{}
\date{}
\vspace{50mm}

\begin{flushright}
\end{flushright}
\vspace{0mm}

\begin{center}
\vspace{10mm}
{\LARGE{Higher order analogues of unitarity condition}}
 \\ \vspace{4mm}
 {\LARGE{for quantum R-matrices}}
\\
\vspace{14mm} {\large {Andrei Zotov} }\\
 \vspace{10mm}
{\small{\sf Steklov Mathematical
Institute  RAS, Gubkina str. 8, Moscow, 119991,  Russia}}\\
\end{center}
\vspace{-8mm}
\begin{center}\footnotesize{{\rm E-mail:}{\rm\ \
 zotov@mi.ras.ru}}\end{center}

 \begin{abstract}
We prove a family of $n$-th order identities for quantum
$R$-matrices of Baxter-Belavin type in fundamental representation.
The set of identities includes the unitarity condition as the
simplest one ($n=2$). Our study is inspired by the fact that the
third order identity provides commutativity of the
Knizhnik-Zamolodchikov-Bernard connections. On the other hand the
same identity gives rise to $R$-matrix valued Lax pairs for the
classical integrable systems of Calogero type. The latter
construction uses interpretation of quantum $R$-matrix as matrix
generalization of the Kronecker function. We present a proof of the
higher order scalar identities for the Kronecker functions which is
then naturally generalized to the $R$-matrix identities.
 \end{abstract}

\newpage

{\small{

\tableofcontents

}}


\section{Introduction and summary}
\setcounter{equation}{0}

Quantum ${\rm GL}(N,\mC)$ $R$-matrix is a solution of the quantum
Yang-Baxter equation
  \beq\label{e01}
  \begin{array}{c}
  \displaystyle{
 R^\hbar_{12}(z_1,z_2)R^\hbar_{13}(z_1,z_3)R^\hbar_{23}(z_2,z_3)
 =R^\hbar_{23}(z_2,z_3)R^\hbar_{13}(z_1,z_3)R^\hbar_{12}(z_1,z_2)\,.
 }
 \end{array}
 \eq
 In fundamental representation $R$-matrix $R_{12}^\hbar$ is an element of $\Mat^{\otimes
 2}$. $R_{ab}^\hbar$ with $1\leq a,b\leq n$ is understood as the element
 of $\Mat^{\otimes n}$ which is identical operator in all components of the tensor product except $a$ and $b$. The
 projection on $a,b$-th components coincide with $R_{12}^\hbar$. The
 $R$-matrices under consideration depend on only difference of the
 spectral parameters. The following notation is used:
  \beq\label{e02}
  \begin{array}{c}
  \displaystyle{
 R_{ab}^\hbar=R^\hbar_{ab}(z_a-z_b)\,.
 }
 \end{array}
 \eq
 We deal with $R$-matrices which include the rational Yang's one \cite{Yang}
  \beq\label{e03}
  \begin{array}{c}
  \displaystyle{
 R^{\rm Yang}_{12}(z_1,z_2)=\frac{1\otimes
1}{\hbar}+\frac{NP_{12}}{z_1-z_2}\,,\ \ \
P_{12}=\sum\limits_{i,j=1}^N E_{ij}\otimes E_{ji}
 }
 \end{array}
 \eq
 as the simplest case, its deformations
 \cite{Cherednik,Smirnov,LOZ8}, trigonometric $R$-matrices
 \cite{Cherednik,Zabr} and elliptic Baxter-Belavin's $R$-matrices
 \cite{Baxter,Belavin}.

 Besides (\ref{e01}) an $R$-matrix satisfies also the unitarity
 condition $R_{12}^\hbar R_{21}^\hbar=1\otimes 1$. We write it using
 different normalization:
 \beq\label{e11}
 \displaystyle{
R^\hbar_{12}
R^\hbar_{21}=N^2\phi(N\hbar,z_1-z_2)\phi(N\hbar,z_2-z_1)\,\,1\otimes
1=N^2 (\wp(N\hbar)-\wp(z_1-z_2))\,\,1\otimes 1\,,
 }
  \eq
where $\phi(\eta,z)$ is the Kronecker function \cite{Weil}.
Depending on a choice of rational, trigonometric or elliptic case it
is equal to\footnote{The notations for elliptic functions in this
paper coincide with those given in \cite{LOZ10} and \cite{LOZ11}
(see Appendix). One can also find in that papers the definition of
the elliptic $R$-matrix satisfying the properties and identities
which are discussed here. Some important definitions and properties
are given in this paper as may be necessary.}
  \beq\label{e04}
  \begin{array}{c}
  \displaystyle{
 \phi(\eta,z)=\left\{
   \begin{array}{l}
    1/\eta+1/z\,,
    \\
    \coth(\eta)+\coth(z)\,,
    \\
    \frac{\vth'(0)\vth(\eta+z)}{\vth(\eta)\vth(z)}
   \end{array}
 \right.
 }
 \end{array}
 \eq
 %
%
The r.h.s. of (\ref{e11}) contains the following function:
  \beq\label{e05}
  \begin{array}{c}
  \displaystyle{
 \wp(z)=\left\{
   \begin{array}{l}
 1/z^2\,,
\\
   1/\sinh^2(z)\,,
\\
    \wp(z)\ -\ \hbox{Weierstrass elliptic $\wp$-function}
   \end{array}
 \right.
 }
 \end{array}
 \eq
In (\ref{e11}) we have already used the identity
  \beq\label{e06}
  \begin{array}{c}
  \displaystyle{
\phi(\eta,z)\phi(\eta,-z)=\wp(\eta)-\wp(z)\,.
 }
 \end{array}
 \eq
It was observed in \cite{Pol} and later in \cite{LOZ9,LOZ10}  that
the Belavin's $R$-matrix can be viewed as matrix generalization of
the elliptic Kronecker function\footnote{Originally, the idea to
consider quantum $R$-matrix as matrix generalization of a scalar
function was proposed in \cite{BazhStrog}. In that papers the
standard normalization of unitarity condition $R_{12}^\hbar
R^\hbar_{21}=1\otimes 1$ was used.}. In particular, it satisfies the
matrix analogue of the Fay identity \cite{Fay,Weil}
  \beq\label{e07}
  \begin{array}{c}
  \displaystyle{
\phi(\hbar,z)\phi(\eta,w)=\phi(\hbar-\eta,z)\phi(\eta,z+w)+\phi(\eta-\hbar,w)\phi(\hbar,z+w)
 }
 \end{array}
 \eq
known as the associative Yang-Baxter equation \cite{Aguiar}:
 %
  \beq\label{e08}
  \begin{array}{c}
  \displaystyle{
 R^\hbar_{ac}
 R^{\eta}_{cb}=R^{\eta}_{ab}R_{ac}^{\hbar-\eta}+R^{\eta-\hbar}_{cb}R^\hbar_{ab}\,.
 }
 \end{array}
 \eq
As a consequence of (\ref{e08}) together with the unitarity
condition (\ref{e11}) and the skew-symmetry\footnote{The property
(\ref{e09}) is the matrix analogue of the Kronecker function
property $\phi(\eta,z)=-\phi(-\eta,-z)$. In \cite{Pol} $R^\hbar(z)$
was treated as the classical $r$-matrix and (\ref{e09}) -- as the
unitarity condition while (\ref{e11}) was not used.}
  \beq\label{e09}
  \begin{array}{c}
  \displaystyle{
 R^\hbar_{ab}(z_a-z_b)=-R_{ba}^{-\hbar}(z_b-z_a)
 }
 \end{array}
 \eq
 one can derive \cite{LOZ9,LOZ11} that $R^\hbar_{ab}$ satisfies the Yang-Baxter equation (\ref{e01})
and the following cubic identity:
  \beq\label{e12}
  \begin{array}{c}
  \displaystyle{
 R^\hbar_{12} R^\hbar_{23} R^\hbar_{31}+R^\hbar_{13}
 R^\hbar_{32} R^\hbar_{21}=-N^3\wp'(N\hbar)\,\,1\otimes 1\otimes
 1=
 }
 \\ \ \\
  \displaystyle{
 =N^3\Big(\phi(N\hbar,z_{12})\phi(N\hbar,z_{23})\phi(N\hbar,z_{31})+
 \phi(N\hbar,z_{13})\phi(N\hbar,z_{32})\phi(N\hbar,z_{21})\Big)\,\,1\otimes
 1\otimes
 1\,,
 }
 \end{array}
 \eq
 where $z_{ab}=z_a-z_b$. Some important applications of the latter identity
 are
 discussed in the end of the paper.

\vskip3mm

\noindent{\bf Purpose of paper} is to generalize the unitarity
condition (\ref{e11}) and identity (\ref{e12}) to its higher order
analogues. The results are summarized in 

\noindent{\bf Theorem} {\em Let quantum $R$-matrix satisfies the
unitary condition (\ref{e11}) and the associative Yang-Baxter
equation (\ref{e08}). Then it also satisfies the following set of
$n$-th order identities for any $n\in\mathbb Z_+$:}
  \beq\label{e14}
  \begin{array}{l}
  \displaystyle{
 \sum\limits_{\hbox{\tiny{$
\begin{array}{c}
1\leq i_1 ... i_{n\!-\!1}\leq n
\\
i_c\neq a;\ i_b\neq i_c
\end{array}
$}}} \! R^\hbar_{ai_1} R^\hbar_{i_1 i_2}\,...\,
R^\hbar_{i_{n-2}i_{n-1}} R^\hbar_{i_{n-1}a}
 =
 }
 \end{array}
 \eq
 $$
  \begin{array}{r}
  \displaystyle{
\hskip0mm =\underbrace{1\otimes...\otimes 1}_{\hbox{\small{\em n
times}}}\, N^n\!
 \sum\limits_{\hbox{\tiny{$
\begin{array}{c}
1\leq i_1 ... i_{n\!-\!1}\leq n
\\
i_c\neq a;\ i_b\neq i_c
\end{array}
$}}}\! \phi({N\hbar},z_{a}-z_{i_1})\phi({N\hbar},z_{i_1}-z_{i_2})...
\phi({N\hbar},z_{i_{n-1}}-z_a) \,,
 }
 \end{array}
 $$
{\em where $a$ is a fixed index $1\leq a \leq n$. For $n\geq 3$
(\ref{e14}) can be rewritten as follows:}
  \beq\label{e15}
  \begin{array}{c}
  \displaystyle{
 \sum\limits_{\hbox{\tiny{$
\begin{array}{c}
1\leq i_1 ... i_{n\!-\!1}\leq n
\\
i_c\neq a;\ i_b\neq i_c
\end{array}
$}}}
 R^\hbar_{ai_1} R^\hbar_{i_1 i_2}\,...\, R^\hbar_{i_{n-2}i_{n-1}} R^\hbar_{i_{n-1}a}=
 \underbrace{1\otimes...\otimes 1}_{\hbox{\small{\em n
times}}}\,(-N)^n\frac{d^{(n-2)}}{d\eta^{(n-2)}}\,\wp(\eta)\left.\right|_{\eta=N\hbar}\,.
 }
 \end{array}
 \eq
The products of $R$-matrices in the sums in (\ref{e14}), (\ref{e15})
contain all possible values of distinct indices from the interval
$1,...,n$. The total number of terms equals $(n-1)!$. Indeed, since
$a$ is fixed the summation index $i_1$ has $n-1$ possible values.
After $i_1$ is also fixed the next index $i_2$ has $n-2$ possible
values $\{1,...,n\}\setminus\{a,i_1\}$, e.t.c. Therefore, the total
number of possible sets of indices equals $(n-1)!$. For example, for
$n=3$
 (\ref{e15}) reproduces (\ref{e12}), and for $n=4$ it reads as follows:
 \beq\label{e13}
 \begin{array}{c}
 \displaystyle{
R^\hbar_{12}R^\hbar_{23}R^\hbar_{34}R^\hbar_{41}+R^\hbar_{12}R^\hbar_{24}R^\hbar_{43}R^\hbar_{31}+
R^\hbar_{13}R^\hbar_{32}R^\hbar_{24}R^\hbar_{41}+R^\hbar_{13}R^\hbar_{34}R^\hbar_{42}R^\hbar_{21}
 }
 \\ \ \\
 \displaystyle{
+
R^\hbar_{14}R^\hbar_{42}R^\hbar_{23}R^\hbar_{31}+R^\hbar_{14}R^\hbar_{4
3}R^\hbar_{32}R^\hbar_{21}=N^4\wp''(N\hbar)\,\,1\otimes 1\otimes
1\otimes 1\,.
 }
 \end{array}
  \eq

The statement of the Theorem implies of course that for $n\geq 3$
the sum of the Kronecker functions products in the r.h.s of
(\ref{e14}) is equal to $(-1)^n\wp^{(n-2)}(N\hbar)$. This statement
looks like it should be well-known. However, we could not find it in
the literature\footnote{Presumably, this type of identities can be
obtained by combining the elliptic Cauchy determinant formulae
proposed by S.N.M. Ruijsenaars \cite{Ruijs1}.}. In the next Section
we propose a simple proof of this identity in a way which is
naturally generalized to $R$-matrix identities proved in Section 3.
The analogue of (\ref{e14}) for $n=1$ is discussed in the end of the
paper.

\section{Higher order identities for Kronecker function}
\setcounter{equation}{0}

In this Section we will prove

\noindent{\bf Proposition} {\em The Kronecker function
$\phi(\eta,z)$ satisfies the following set of $n$-th order
identities for any $n>2$:
 }
  \beq\label{e21}
  \begin{array}{c}
  \displaystyle{
\sum\limits_{\hbox{\tiny{$
\begin{array}{c}
1\leq i_1 ... i_{n\!-\!1}\leq n
\\
i_c\neq a;\ i_b\neq i_c
\end{array}
$}}}\! \phi(\eta,z_{a}-z_{i_1})\phi(\eta,z_{i_1}-z_{i_2})...
\phi(\eta,z_{i_{n-1}}-z_a)
=(-1)^n\frac{d^{(n-2)}}{d\eta^{(n-2)}}\,\wp(\eta)\,.
 }
 \end{array}
 \eq
Before we proceed further let us introduce one more function
important for our purposes. It is the first Eisenstein
 function\footnote{It is simply related to the Weierstrass zeta-function: $E_1(z)=\zeta(z)+\frac{z}{3}\frac{\vth'''(0)}{\vth'(0)}$.}:
  \beq\label{e091}
  \begin{array}{c}
  \displaystyle{
 E_1(z)=\left\{
   \begin{array}{l}
 1/z\,,
\\
   \coth(z)\,,
\\
    \vth'(z)/\vth(z)
   \end{array}
 \right.\hskip20mm E_1(-z)=-E_1(z)\,.
 }
 \end{array}
 \eq
 It appears in the expansion of the Kronecker function near the pole
 $\eta=0$:
  \beq\label{e095}
  \begin{array}{c}
  \displaystyle{
 \phi(\eta,z)=\eta^{-1}+E_1(z)+\eta\,
 (E_1^2(z)-\wp(z))/2+O(\eta)\,,
 }
 \end{array}
 \eq
in the formula for derivative
  \beq\label{e096}
  \begin{array}{c}
  \displaystyle{
 \frac{d}{d\eta}\,\phi(\eta,z)=(E_1(\eta+z)-E_1(\eta))\phi(\eta,z)
 }
 \end{array}
 \eq
and in degenrated ($\eta=\hbar$) Fay identity (\ref{e07})
 \beq\label{e097}
  \begin{array}{c}
  \displaystyle{
 \phi(\eta,z)\phi(\eta,w)=\phi(\eta,z+w)(E_1(\eta)+E_1(z)+E_1(w)-E_1(z+w+\eta))\,.
 }
 \end{array}
 \eq
Combining (\ref{e096}) and (\ref{e097}) we get
 \beq\label{e24}
  \begin{array}{c}
  \displaystyle{
\frac{d}{d\eta}\,\phi(\eta,z)\equiv\phi'(\eta,z)=(E_1(z+y)-E_1(y))\phi(\eta,z)-\phi(\eta,z+y)\phi(\eta,-y)\,.
 }
 \end{array}
 \eq
Notice that (\ref{e24}) is valid for any $y\in\mathbb C$ while its
l.h.s is independent of $y$.

A straightforward way for proving of (\ref{e21}) type relations is
to compare the poles and residues of both sides. It is easy to show
that the l.h.s is a double-periodic function of $\eta$ and
$z_1,...,z_n$. Moreover, one can verify that the residues at
$z_i=z_j$ equal zero. However, this only means that the l.h.s is a
double-periodic function of $\eta$ which behaves as $1/\eta^n$ near
$\eta=0$. In this way it is hard to fix it, i.e. to prove the
absence (in the r.h.s.) of terms
$c_k(\tau)\frac{d^{k}}{d\eta^{k}}\wp(\eta)$ with $k<n-2$. Instead,
we suggest another proof of (\ref{e21}) based on  (\ref{e24}).

\vskip3mm\underline{\em{Proof of Proposition}}:\vskip3mm

\noindent The proof of (\ref{e21}) is by induction on $n$. For $n=3$
 $$
\phi(\eta,z_{1}-z_{2})\phi(\eta,z_{2}-z_{3})\phi(\eta,z_{3}-z_1)+
\phi(\eta,z_{1}-z_{3})\phi(\eta,z_{3}-z_{2})\phi(\eta,z_{2}-z_1) =
-\wp'(\eta)
 $$
 it is well-known. On one can verify it directly by comparing the structure of poles and
 residues. Alternatively, one can use (\ref{e097}) and then (\ref{e06}) to rewritten it in a more recognizable
 form:
 $$
 E_1(\eta+z)+E_1(\eta-z)-2E_1(\eta)=\frac{\wp'(\eta)}{\wp(\eta)-\wp(z)}\,,
 \ z=z_a-z_b\,.
 $$
Let (\ref{e21}) is true for $n$. Differentiate its both sides with
respect to $\eta$. The r.h.s. is then equal to
  \beq\label{e221}
  \begin{array}{c}
  \displaystyle{
-(-1)^{n+1}\frac{d^{(n-1)}}{d\eta^{(n-1)}}\,\wp(\eta)\,,
 }
 \end{array}
 \eq
that is the r.h.s. of (\ref{e21}) taken for $n:=n+1$ with the
opposite sign.

 For the l.h.s we have:
  \beq\label{e23}
  \begin{array}{c}
  \displaystyle{
\sum\limits_{\hbox{\tiny{$
\begin{array}{c}
1\leq i_1 ... i_{n\!-\!1}\leq n
\\
i_c\neq a;\ i_b\neq i_c
\end{array}
$}}}\Big( \phi'_{a i_1}(\eta)\phi_{i_1i_2}(\eta)...
\phi_{i_{n\!-\!1}a}(\eta)
+...+\phi_{a i_1}(\eta)\phi_{i_1i_2}(\eta)...
\phi'_{i_{n\!-\!1}a}(\eta) \Big)\,,
 }
 \end{array}
 \eq
 where the short notations
 $\phi_{ij}(\eta)=\phi(\eta,z_{i}-z_{j})$ and
 $\phi'_{ij}(\eta)=\frac{d}{d\eta}\,\phi(\eta,z_{i}-z_{j})$
 are used.


   Substitute (\ref{e24}) into (\ref{e23})
choosing $y$ each time as follows: for $\phi'_{ij}(\eta)$ let
  \beq\label{e222}
  \begin{array}{c}
  \displaystyle{
y=z_{j}-z_{n+1}\,.
 }
 \end{array}
 \eq
 The answer for (\ref{e23}) after the substitution consists of terms of two types:

1. of the form $\phi^n E_1$;

2. of the form $\phi^{n+1}$.

The first type terms cancel out for each expression inside the
brackets in (\ref{e23}) due to skew-symmetry (\ref{e091}) of $E_1$
function. Indeed, summing up all such terms with fixed values of
$i_1,..., i_{n-1}$ we obtain:
  \beq\label{e25}
  \begin{array}{c}
  \displaystyle{
0= \phi_{a i_1}(\eta)\phi_{i_1i_2}(\eta)...
\phi_{i_{n\!-\!1}a}(\eta)\Big(E_1(z_a-z_{i_1}+z_{i_1}-z_{n+1})-E_1(z_{i_1}-z_{n+1})+
 }
  \end{array}
 \eq
 $$
  \displaystyle{
E_1(z_{i_1}\!-\!z_{i_2}\!+\!z_{i_2}\!-\!z_{n+1})\!-\!E_1(z_{i_2}\!-\!z_{n+1})+\!...\!+E_1(z_{i_{n\!-\!1}}\!-\!z_a+z_a\!-\!z_{n+1})
\!-\!E_1(z_a\!-\!z_{n+1}) \Big)\,.
 }
 $$

The total sum of the second type terms (of from $\phi^{n+1}$) is
exactly the l.h.s. of (\ref{e21}) taken for $n:=n+1$ with the common
opposite sign:
  \beq\label{e231}
  \begin{array}{l}
  \displaystyle{
\sum\limits_{\hbox{\tiny{$
\begin{array}{c}
1\leq i_1 ... i_{n-1}\leq n
\\
i_c\neq a;\ i_b\neq i_c
\end{array}
$}}}\Big( \phi'_{a i_1}(\eta)\phi_{i_1i_2}(\eta)...
\phi_{i_{n-1}a}(\eta)
+...+\phi_{a i_1}(\eta)\phi_{i_1i_2}(\eta)... \phi'_{i_{n-1}a}(\eta)
\Big)=
 }
 \\ \ \\
   \displaystyle{
-\sum\limits_{\hbox{\tiny{$
\begin{array}{c}
1\leq i_1 ... i_{n-1}\leq n
\\
i_c\neq a;\ i_b\neq i_c
\end{array}
$}}}\Big( \phi_{a, n+1}(\eta)\phi_{n+1,
i_1}(\eta)\phi_{i_1i_2}(\eta)... \phi_{i_{n-1}a}(\eta)+...
 }
 \\ \ \\
  \displaystyle{
+\,\phi_{a i_1}(\eta)\phi_{i_1i_2}(\eta)...
\phi_{i_{n-1},n+1}(\eta)\phi_{n+1,a}(\eta)
\Big)=-\sum\limits_{\hbox{\tiny{$
\begin{array}{c}
1\leq i_1 ... i_{n}\leq n
\\
i_c\neq a;\ i_b\neq i_c
\end{array}
$}}} \phi_{a i_1}(\eta)\phi_{i_1i_2}(\eta)... \phi_{i_{n}a}(\eta)\,.
 }
 \end{array}
 \eq
 The last equality is easily verified by considering the structure
of summation indices. All of them are distinct for each term and
differ from the "outer" one $a$.\footnote{The value of the "outer"
index $a$ is in fact not important here: see remark (\ref{e22}).}
The same expression is obtained from the second type terms after the
substitution. At the same time the total number of the second type
terms in (\ref{e23}) (and therefore in (\ref{e231})) equals $n!$
(before differentiating with respect to $\eta$ we had $(n-1)!$
terms\footnote{See the comment after (\ref{e15}).} and then applied
the Leibniz rule to each term) which coincide with the number of
terms in (\ref{e21}) taken for $n:=n+1$.

Formula (\ref{e231}) together with (\ref{e221}) finishes the
inductive proof. $\blacksquare$

\section{Higher order R-matrix identities}
\setcounter{equation}{0}

In this Section we prove the Theorem (\ref{e14}), (\ref{e15}). The
proof is similar to the one of the Proposition from the previous
Section.

The $R$-matrix analogue of the first Eisenstein function
(\ref{e091}) is the classical $r$-matrix $r_{ab}(z)$. It appears in
the classical limit in the same way as $E_1$ in (\ref{e095}):
  \beq\label{e41}
  \begin{array}{c}
  \displaystyle{
R_{12}^\hbar(z)=\frac{1}{\hbar}\,1\otimes 1
+r_{12}(z)+\hbar\,m_{12}(z)+O(\hbar)\,.
 }
 \end{array}
 \eq
Similarly to $E_1$-function the classical $r$-matrix is
skew-symmetric\footnote{The latter follows from either unitarity
condition (in $\hbar^{-1}$, $\hbar^0$ orders of expansion) or from
the skew-symmetry (\ref{e09}) (in $\hbar^{0}$, $\hbar^1$ orders of
expansion).}:
  \beq\label{e42}
  \begin{array}{c}
  \displaystyle{
r_{ab}(z_a-z_b)=-r_{ba}(z_b-z_a)\,,\ \ \ \
m_{ab}=m_{ba}=\frac{1}{2}\,\left(r_{ab}^2- 1\otimes
1\,N^2\wp(z_a-z_b)\right)\,.
 }
 \end{array}
 \eq

The $R$-matrix analogue of (\ref{e24}) comes from the associative
Yang-Baxter equation (\ref{e08}) and (\ref{e41}) in the limit
$\eta\rightarrow \hbar\,$:
  \beq\label{e43}
  \begin{array}{c}
  \displaystyle{
 \p_\hbar R^\hbar_{ab}\equiv J_{ab}^\hbar
 =R^{\hbar}_{ab}r_{ac}+r_{cb}R^\hbar_{ab}-R^\hbar_{ac}R^\hbar_{cb}\,.
 }
 \end{array}
 \eq

\vskip3mm\underline{\em{Proof of Theorem (\ref{e14}),
(\ref{e15})}}:\vskip3mm

\noindent The statement (\ref{e14}) for $n=2$ is the unitarity
condition (\ref{e11}), for $n=3$ -- it is the known from
\cite{LOZ9,LOZ11} identity (\ref{e12}), and for $n=1$ it is due to
(\ref{e55}). Let us prove (\ref{e15}). Then (\ref{e14}) follows from
the scalar analogue of (\ref{e15}) -- Proposition (\ref{e21}).

The idea of the proof repeats the one given for (\ref{e21}).
The proof of (\ref{e15}) is by induction on $n$. Let it is true for
$n$. Multiply both sides of (\ref{e15}) by $\otimes 1$ ($1$ is
identity $N\times N$ matrix) and differentiate them with respect to
$\hbar$. Then the r.h.s. becomes equal to
  \beq\label{e35}
  \begin{array}{c}
  \displaystyle{
-\underbrace{1\otimes...\otimes 1}_{\hbox{\small{\em $n+1$
times}}}\,(-N)^{n+1}\frac{d^{(n-1)}}{d\eta^{(n-1)}}\,\wp(\eta)\left.\right|_{\eta=N\hbar}\,,
 }
 \end{array}
 \eq
that is the r.h.s. of (\ref{e15}) taken for $n:=n+1$ with the
opposite sign.

The l.h.s becomes
  \beq\label{e36}
  \begin{array}{c}
  \displaystyle{
 \sum\limits_{\hbox{\tiny{$
\begin{array}{c}
1\leq i_1 ... i_{n\!-\!1}\leq n
\\
i_c\neq a;\ i_b\neq i_c
\end{array}
$}}}
 \Big( J^\hbar_{ai_1} R^\hbar_{i_1 i_2}\,...\, R^\hbar_{i_{n-2}i_{n-1}} R^\hbar_{i_{n-1}a} +...+
 R^\hbar_{ai_1} R^\hbar_{i_1 i_2}\,...\, R^\hbar_{i_{n-2}i_{n-1}} J^\hbar_{i_{n-1}a} \Big)\,.
 }
 \end{array}
 \eq
The latter is analogues to (\ref{e23}). As in the scalar case,
plugging $J^\hbar_{ab}$ from (\ref{e43}) with $c=n+1$ we obtain the
l.h.s. of (\ref{e15}) taken for $n:=n+1$ with the common opposite
sign. This answer comes from $-R_{i,n+1}^\hbar R_{n+1,j}$ type terms
in (\ref{e43}) as in the proof of the sclar identities. Let us only
comment the cancellation of terms containing classical $r$-matrices
$r_{ab}$. The input of such terms into the expression inside the
brackets in (\ref{e36}) equals
  \beq\label{e37}
  \begin{array}{c}
  \displaystyle{
(R^\hbar_{ai_1}r_{a,n+1}+r_{n+1,i_1}R^\hbar_{ai_1} ) R^\hbar_{i_1
i_2}\,...\,  R^\hbar_{i_{n-1}a} +
 R^\hbar_{ai_1}(R^\hbar_{i_1
i_2}r_{i_1,n+1}+r_{n+1,i_2}R^\hbar_{i_1 i_2})R^\hbar_{i_2i_3}\,...\,
R^\hbar_{i_{n-1}a}+
 }
\\ \ \\
  \displaystyle{
...+
 R^\hbar_{ai_1} R^\hbar_{i_1 i_2}\,...\, R^\hbar_{i_{n-2}i_{n-1}}
  (R^\hbar_{i_{n-1}a} r_{i_{n-1},n+1} +r_{n+1,a}R^\hbar_{i_{n-1}a} )
 }
 \end{array}
 \eq
Two terms containing $r_{a,n+1}$ are the first and the last ones.
They are cancelled because $r_{a,n+1}$ in the first term can be
moved\footnote{The product $R^\hbar_{i_1 i_2}\,...\,
R^\hbar_{i_{n-2},i_{n-1}}$ does not contain indices $a$ or $n+1$,
and therefore commutes with $r_{a,n+1}$.} to the last but one
position and due to skew-symmetry (\ref{e42}). All other terms are
combined into the commutator
 $$
[r_{n+1,i_1}+r_{n+1,i_2}+...+r_{n+1,i_{n-1}},R^\hbar_{ai_1}
R^\hbar_{i_1 i_2}\,...\,  R^\hbar_{i_{n-1}a}]
 $$
 because all $r$-matrices with indices $r_{n+1,k}$ can be moved to
 the left positions in the corresponding products, and $r$-matrices with indices
 $r_{k,n+1}$ can be moved to
 the right position in their products. The obtained commutator equals
 zero since the product of $R$-matrices is a scalar operator
by the inductive assumption. Therefore, expression (\ref{e37})
equals zero as well. $\blacksquare$

In the end of the Section let us also write down (\ref{e15}) for
$n=5$:
 \beq\label{e134}
 \begin{array}{c}
 \displaystyle{
 R^\hbar_{15}R^\hbar_{54}R^\hbar_{43}R^\hbar_{32}R^\hbar_{21}+
 R^\hbar_{14}R^\hbar_{45}R^\hbar_{53}R^\hbar_{32}R^\hbar_{21}+R^\hbar_{13}R^\hbar_{35}R^\hbar_{54}R^\hbar_{42}R^\hbar_{21}+
 R^\hbar_{15}R^\hbar_{53}R^\hbar_{34}R^\hbar_{42}R^\hbar_{21}
 }
 \\ \ \\
 \displaystyle{
 +R^\hbar_{13}R^\hbar_{34}R^\hbar_{45}R^\hbar_{52}R^\hbar_{21}+R^\hbar_{14}R^\hbar_{43}R^\hbar_{35}R^\hbar_{52}R^\hbar_{21}+
 R^\hbar_{12}R^\hbar_{25}R^\hbar_{54}R^\hbar_{43}R^\hbar_{31}+
 R^\hbar_{12}R^\hbar_{24}R^\hbar_{45}R^\hbar_{53}R^\hbar_{31}
 }
 \\ \ \\
 \displaystyle{
 +R^\hbar_{15}R^\hbar_{54}R^\hbar_{42}R^\hbar_{23}R^\hbar_{31}+
 R^\hbar_{14}R^\hbar_{42}R^\hbar_{25}R^\hbar_{53}R^\hbar_{31}+
 R^\hbar_{14}R^\hbar_{45}R^\hbar_{52}R^\hbar_{23}R^\hbar_{31}+R^\hbar_{15}R^\hbar_{52}R^\hbar_{24}R^\hbar_{43}R^\hbar_{31}
 }
 \\ \ \\
 \displaystyle{
 +
 R^\hbar_{12}R^\hbar_{23}R^\hbar_{35}R^\hbar_{54}R^\hbar_{41}+
 R^\hbar_{12}R^\hbar_{25}R^\hbar_{53}R^\hbar_{34}R^\hbar_{41}+R^\hbar_{13}R^\hbar_{32}R^\hbar_{25}R^\hbar_{54}R^\hbar_{41}+
 R^\hbar_{15}R^\hbar_{53}R^\hbar_{32}R^\hbar_{24}R^\hbar_{41}
 }
 \\ \ \\
 \displaystyle{
 +
 R^\hbar_{13}R^\hbar_{35}R^\hbar_{52}R^\hbar_{24}R^\hbar_{41}+R^\hbar_{15}R^\hbar_{52}R^\hbar_{23}R^\hbar_{34}R^\hbar_{41}+
 R^\hbar_{12}R^\hbar_{23}R^\hbar_{34}R^\hbar_{45}R^\hbar_{51}+
 R^\hbar_{12}R^\hbar_{24}R^\hbar_{43}R^\hbar_{35}R^\hbar_{51}
 }
 \\ \ \\
 \displaystyle{
 +R^\hbar_{13}R^\hbar_{32}R^\hbar_{24}R^\hbar_{45}R^\hbar_{51}+
 R^\hbar_{14}R^\hbar_{43}R^\hbar_{32}R^\hbar_{25}R^\hbar_{51}+
 R^\hbar_{13}R^\hbar_{34}R^\hbar_{42}R^\hbar_{25}R^\hbar_{51}+R^\hbar_{14}R^\hbar_{42}R^\hbar_{23}R^\hbar_{35}R^\hbar_{51}
 }
 \\ \ \\
 \displaystyle{
=-N^5\wp'''(N\hbar)\,\,1\otimes 1\otimes 1\otimes 1\otimes 1\,.
 }
 \end{array}
  \eq

\section{Applications and remarks}
\setcounter{equation}{0}

\begin{itemize}

\item The permutation group of the set $(z_1,...,z_n)$ transforms an
$R$-matrix identities (\ref{e14})-(\ref{e15}) with some "outer"
index $a$ to identity of the same form for different $\ti a\neq a$.
One should rename a set of $\{z_k\}$ and act on
(\ref{e14})-(\ref{e15}) by the corresponding product of permutation
operators (from both sides). In the scalar case the identity
(\ref{e21}) is invariant with respect to the action of the
permutation group: its l.h.s is in fact independent of index $a$
because one can rearrange the products of functions $\phi$ in a way
that the total sum acquires the form of the same identity for a
different value of the index $\ti a \neq a$.
For example, for $n=3$ obviously
  \beq\label{e22}
  \begin{array}{c}
  \displaystyle{
\phi(\eta,z_{1}-z_{2})\phi(\eta,z_{2}-z_{3})\phi(\eta,z_{3}-z_1)+
\phi(\eta,z_{1}-z_{3})\phi(\eta,z_{3}-z_{2})\phi(\eta,z_{2}-z_1) = }
\\ \ \\
  \displaystyle{
\phi(\eta,z_{2}-z_{1})\phi(\eta,z_{1}-z_{3})\phi(\eta,z_{3}-z_2)+
\phi(\eta,z_{2}-z_{3})\phi(\eta,z_{3}-z_{1})\phi(\eta,z_{1}-z_2)
 }
 \end{array}
 \eq
Here $a=1$ and $\ti a=2$. The same holds true for any $n$, $a$, $\ti
a$.

\item The $R$-matrix identity (\ref{e14}) makes some sense even for
$n=1$: for $R_{ab}^\hbar(z)$ we can identify the $a$-th and $b$-th
tensor components keeping $z\neq 0$. Then
  \beq\label{e55}
  \begin{array}{c}
  \displaystyle{
 R_{aa}^\hbar(z)=1_a\, N\phi(N\hbar,\frac{z}{N})\,,
 }
 \end{array}
 \eq
 where $1_a$ is the identity $N\times N$ matrix in the $a$-th
 component. To prove (\ref{e55}) consider the most general -
 elliptic Belavin's $R$-matrix \cite{Belavin}. In our notations and
 with our normalization it acquires the form:
 \beq\label{e56}
 \begin{array}{c}
  \displaystyle{
R^\hbar_{12}(z)=\sum\limits_{\al\in\, {\mathbb Z}_N\times {\mathbb
Z}_N}\vf_\al(z,\om_\al+\hbar)\,T_\al\otimes T_{-\al}\in
\hbox{Mat}(N,\mathbb C)^{\otimes 2}\,,
 }
 \end{array}
 \eq
 where
 \beq\label{e57}
 \begin{array}{c}
  \displaystyle{
\vf_\al(z,\om_\al+\hbar)=\exp(2\pi\imath
\frac{\al_2}{N}z)\,\phi(z,\om_\al+\hbar)\,,\ \ \
\om_\al=\frac{\al_1+\al_2\tau}{N}\,,
 }
 \end{array}
 \eq
while $\{T_\al\}$ is a special basis in $\hbox{Mat}(N,\mathbb C)$:
  \beq\label{e58}
 \begin{array}{c}
  \displaystyle{
T_\al T_\be=\kappa_{\al,\be} T_{\al+\be}\,,\ \ \
\kappa_{\al,\be}=\exp\left(\frac{\pi \imath}{N}(\be_1
\al_2-\be_2\al_1)\right)\,,
 }
 \end{array}
 \eq
i.e.
  \beq\label{e59}
 \begin{array}{c}
  \displaystyle{
T_\al T_{-\al}=1_{N\times N}\,.
 }
 \end{array}
 \eq
See details in e.g. Appendix of \cite{LOZ11}. From (\ref{e56}) and
(\ref{e59}) we conclude that
 \beq\label{e65}
 \begin{array}{c}
  \displaystyle{
R^\hbar_{11}(z)=1_{N\times N}\sum\limits_{\al\in\, {\mathbb
Z}_N\times {\mathbb Z}_N}\vf_\al(z,\om_\al+\hbar)\,.
 }
 \end{array}
 \eq
The latter sum in the l.h.s. is equal to $N\phi(N\hbar,z/N)$. This
is particular case of the finite Fourier transformation formulae for
the Kronecker function \cite{LOZ10}.

\item The quantum $R$-matrices can be used for construction of the
classical Lax pairs for the Calogero type models \cite{LOZ9}. The
${\rm gl}_n$ Lax matrix is of the form:
  \beq\label{e67}
 \begin{array}{c}
  \displaystyle{
\mathcal L(\hbar)=\sum\limits_{a,b=1}^{n} \ti{\mathrm E}_{ab}\otimes
\mathcal L_{ab}(\hbar)\,,\ \ \ \mathcal
L_{ab}(\hbar)=\delta_{ab}p_a\,1_a\otimes
1_b+\nu(1-\delta_{ab})R_{ab}^\hbar(z_a-z_b)\,,
 }
 \end{array}
 \eq
where $\ti{\mathrm E}_{ab}$ is the standard basis in ${\rm
Mat}(n,\mC)$, the set of $R$-matrix spectral parameters $\{z_a\}$ --
are the Calogero particles coordinates, $p_a$ - particles momenta,
and $\hbar$ plays the role of the spectral parameter. Formula
(\ref{e67}) generalizes the Krichever's answer \cite{Krich1}. The
following guess was made in \cite{LOZ9}: the diagonal elements
(blocks) of $\tr \mathcal L^k(\hbar)$ are the scalar operators with
the coefficient equals to the same element computed for $\tr
l(\hbar)^k$ -- the (usual) Krichever's Lax matrix
$l_{ab}(\hbar)=p_a\delta_{ab}+\nu
(1-\delta_{ab})N\phi(N\hbar,z_a-z_b)$. For $n=2$ this statement
follows from the unitarity condition (\ref{e11}) and for $n=3$ it
follows from the cubic identity (\ref{e12}). Obviously the statement
of the Theorem (\ref{e14}) is of the same type. In fact, (\ref{e14})
can be considered as a part of the guess for $k=n$.

\item It follows from the quantum Yang-Baxter equation (\ref{e01})
and the cubic identity (\ref{e12}) that
 \beq\label{e70}
 \begin{array}{c}
  \displaystyle{
 [r_{ab},m_{ac}+m_{bc}]+[r_{ac},m_{ab}+m_{bc}]=0\,,
 }
 \end{array}
 \eq
where $r_{ab}$ and $m_{ab}$ are defined from the classical limit
expansion (\ref{e41}) (see details in \cite{LOZ9}). The latter
equation guarantees the commutativity of different type KZB
connections
 \beq\label{e80}
 \begin{array}{l}
  \displaystyle{
\nabla_a=\p_{z_a}+\sum\limits_{b:b\neq a} r_{ab}(z_a-z_b)\,,\ \
a=1,...,n;
 }
 \\
  \displaystyle{
\nabla_\tau=\p_\tau+\sum\limits_{b>c} m_{bc}(z_b-z_c)\,,
 }
 \end{array}
 \eq
 i.e.  $[\nabla_a,\nabla_\tau]=0$ (the commutativity $[\nabla_a,\nabla_b]=0$ follows
 from the classical Yang-Baxter equation). The obtained identities
 (\ref{e14}), (\ref{e15})
 provide a set of equations for $r_{ab}$, $m_{ab}$ and higher
 coefficients of the expansion (\ref{e41}). The simplest one is the
 $1/\hbar^{n-2}$ order terms in (\ref{e15}):
  \beq\label{e85}
 \begin{array}{c}
  \displaystyle{
\sum\limits_{a,b,c:\ c<a<b}\,
[r_{ca},r_{ab}]_++[r_{ab},r_{bc}]_++[r_{bc},r_{ca}]_+
=-(n-2)\,\sum\limits_{b,c:\ b\neq c} m_{bc}\,.
 }
 \end{array}
 \eq



\end{itemize}

\subsubsection*{Acknowledgments}
The work was performed at the Steklov Mathematical Institute RAS,
Moscow. It was supported by Russian Science Foundation (RSCF) grant
14-50-00005.

\begin{small}

\end{small}

\end{document}